\documentclass[a4paper,12pt]{article}
\usepackage{graphicx, graphics}
\usepackage{latexsym}
\usepackage{cite}
\usepackage{epstopdf}
\usepackage{amssymb}
\usepackage{epsfig}
\usepackage{amssymb}
\author{Parviz Goodarzi \footnote{parviz.goodarzi@abru.ac.ir}
\\ {\small Department of Physics, Ayatollah Boroujerdi University, Boroujerd, Iran}}
\title {Gravitational baryogenesis in non-minimal kinetic coupling model}
\begin{document}

\maketitle
\begin{abstract}
In this work, we consider the gravitational baryogenesis in the framework of non-minimal derivative coupling model.
A mechanism to generate the baryon asymmetry based on the coupling between the derivative of the Ricci scalar curvature
 and the baryon current in context of non-minimal derivative coupling model is investigated.
We show that, in this model, the temperature increases during the reheating periods to the end of reheating period or beginning of radiation dominated era. Therefore the reheating temperature is larger then decoupling temperature.
It can be demonstrated that, the evaluation of baryon asymmetry is not depends on coupling constant.
In this model we can generate baryon asymmetry at low and high reheating temperature, by considering the high friction constraint.
\end{abstract}

\section{Introduction}

One of the greatest puzzles in the standard model of cosmology and astroparticle physics is the dominance of matter against anti-matter in the Universe.
In the other words the number of baryons in the Universe is larger then the number of antibaryons.
The cosmic microwave background radiation \cite{WMAP}, the abundance of the primordial light elements emission from matter-antimatter annihilation \cite{Cohen} and the big bang nucleosynthesis (BBN) \cite{BBN}, indicate dominance of matter over antimatter in the Universe.
The ratio of baryon number density $n_B$ to the entropy density $s$ is observationally obtained from the highly precis measurement of the cosmic microwave background radiation as \cite{Planck 2015}
\begin{equation}\label{1}
Y_B\equiv{n_B\over s}=(0.864\pm 0.016)\times 10^{-10}.
\end{equation}

Sakharov in Ref. \cite{Sakharov} showed that baryon asymmetry may be dynamically generated from the following conditions:
(i) processes that violate baryon number;
(ii) violation of charge (C) and charge parity (CP) symmetry;
(iii) out of the thermal equilibrium.
This three assumptions are now known as Sakharov conditions.

Several interesting and applicable mechanisms for generation of baryon asymmetry which satisfy the above conditions have been proposed.
The first suggestion for this topic used on the out of equilibrium decay of a massive particle
such as a super heavy GUT gauge of Higgs boson where dubbed GUT baryogenesis \cite{Weinberg}.
The other mechanism involving the decay of flat directions in super symmetric models is
known as the Affleck-Dine scenario \cite{Affleck}.
Also the possibility of generating the baryon asymmetry at the electro-weak scale has been considered, these interactions conserve
the sum of baryon and lepton number, which is converted to a baryon asymmetry at the electro-weak scale.
This mechanism is known as lepto-baryogenesis \cite{Fukugita}.
The spontaneous baryogenesis has been proposed with the characteristic generation of baryon asymmetry in thermal equilibrium without the necessity of $C$
and $CP$ violation \cite{Cohen1,Cohen2}.

Davoudiasl et al. in Supergravity have proposed a mechanism for generating
 of the baryon asymmetry on the basis of spontaneous baryogenesis during the expansion of the Universe (dubbed gravitational baryogensis) \cite{Davoudiasl}.
In this approach they introduced an interaction between derivative of the Ricci scalar curvature
and baryon current $J^\mu\partial_\mu R $ which dynamically violate CPT in an expanding Universe.
This coupling in the matter dominated universe and kinetic energy of scalar field dominated it has no problem to generate correct magnitude of baryon asymmetry but in the radiation dominated Universe ($\omega=1/3$) it needs to take into account interaction among massless particles.

During the last years, other scenarios to extend this coupling, has been receiving a great amount of attention from the many authors.
In \cite{Sadjadi} the effect of time dependence of the equation of state parameter on gravitational baryogenesis has been considered.

Gravitational baryogenesis in modified $f(R)$ gravity has been considered in \cite{Lambiase,Ramos}.
Modified gravity theories \cite{Nojiri,Capozziello} can easily avoid problem of generation of baryon asymmetry during radiation dominated Universe, as the ensuing modified equations of motion can lead to very different relations between the scalar curvature and temperature.
Also gravitational baryogenesis in context of $f(T)$ theory of gravity \cite{Capozziello1} with the coupling $J^\mu\partial_\mu T$ has been investigated where $T$ is the torsion scalar \cite{Saridakis}.
Some variant forms of gravitational baryogenesis containing the partial derivative of Gauss-Bonnet scalar $\mathcal{G}$ coupled to baryon current $J^\mu\partial_\mu \mathcal{G}$ are investigated in \cite{Odintsov}. In the case of Gauss-Bonnet baryogenesise the ratio of baryon number density to entropy is non-zero in the radiation dominated Universe.

Gravitational baryogenesis with the novel term $J^\mu\partial_\mu \mathcal{T}$ in minimal and non-minimal $f(R,\mathcal{T})$ gravity (where $\mathcal{T}$ denotes the trace of energy-momentum tensor) have been examined in \cite{Baffou,Sahoo} respectively. They found that interaction proportional to $\partial_\mu \mathcal{T}$ produced unphysical results.

Baryogenesis in $f(Q,\mathcal{T})$ gravity (where $Q$ denotes the nonmetricity) with the coupling $J^\mu\partial_\mu Q$ has been explored and found results that are consistent with observation \cite{Bhattacharjee}.

Generalized gravitational baryogenesis in the frameworks of $f(T,T_G)$ and $f(T,B)$ (where $T_G$ is the teleparallel equivalent to the Gauss-Bonnet term, $B=2\nabla_\mu T^\mu$ denotes boundary term between torsion and Ricci scalar) are discussed in \cite{Azhar,Nozari}. The authors of these papers have chosen power-law form of scale factor $a(t)=m_0t^\gamma$ for each models, and constructed baryon to entropy ratio by assuming that the universe filled by perfect fluid and dark energy.

In \cite{Lima} the baryon asymmetry is generated dynamically during an inflationary epoch powered by ultra-relativistic particle production.
In \cite{Saaidi,Saaidi2,Fukushima} has been suggested that anisotropy of the universe can enhance the generation of the baryon asymmetry.
The authors of \cite{Fukushima} clarified, if we into account the gravitino problem $(T_{RD}<10^9GeV)$, gravitational baryogenesis \cite{Davoudiasl} is incapable to explain generation of sufficient baryon asymmetry. They show that if there exists a huge shear in the radiation dominated era there is a little possibility of gravitational baryogenesis.

During the last years, the non-minimal derivative coupling model has been receiving a great amount of attention from the community of theoreticians for explanation of inflation, structure formations, warm inflation and dark energy \cite{Germani,Germani2,Germani3}. 
Therefore, we are interested in investigating the gravitational baryogenesis mechanism in the framework of non-minimal derivative coupling model.
  
Non-minimal derivative coupling $G_{\mu\nu}\partial^\mu\varphi\partial^\nu\varphi$ is one of the operators of Horndeski’s scalar tensor theory \cite{Horndeski}. Horndeski’s theory is the most general scalar-tensor theory having second-order field equations in four dimensions \cite{Tsutomu}.
Extended scalar-tensor theories of gravity can be obtained by generalised conformal or disformal transformations of Horndeski and beyond Horndeski actions \cite{Nojiri,Nojiri2,Gao}. Lagrangian for Horndeski’s theory contains several operators, where non-minimal derivative coupling has received more attention among them.

Non-minimal derivative coupling primordially in \cite{Amendola,Charmousis}, without introducing any degree of freedom more then the inflaton and the massless graviton, violation of quantum gravity bound and violation of unitarity bounds \cite{Germani3}, used to explain Higgs inflation \cite{Germani}.

An emphasize feature of the non-minimal derivative coupling model with the Einstein tensor is that the mechanism of the gravitationally enhanced friction during inflation, by which even too steep potentials with theoretically natural model parameters can drive cosmic acceleration \cite{Germani2,Germani3}.
In particular, this leads to the range of parameters in which inflationary attractors exist is greatly expanded.

The reheating process after the slow roll inflation in the non-minimal derivative coupling model \cite{sadjadi,sadjadi1,sadjadi2,sadjadi3,nozari} show that, inflaton begins a coherent rapid oscillation. At the end of slow roll inflation, during reheating period, inflaton decays to radiation and reheats the Universe.
The temperature of the Universe increases to the end of scalar field dominated in contrast to the standard reheating in general relativity.
This different behavior of the non-minimal derivative coupling model during the reheating period makes us interested in investigating the gravitational baryogenesis in this model.

Thus it is well motivated to propose the gravitational baryogenesis in context of non-minimal derivative coupling model.
In fact, the purpose of the current work is to investigate gravitational baryogenesis during reheating and radiation dominated universe in this model and compare results with the observation.
Moreover, the alternative description of the current work, is to consider the effect of “high friction constraint” on the baryon to entropy ratio and the values of the model parameters. We would like to investigate the possibility of baryon asymmetry in the presence of high friction condition in this model.

We shall demonstrate here, in contrast to the standard gravitation baryogenesis where could not explain baryon asymmetry in low reheating temperature \cite{Davoudiasl}, in the non-minimal derivative coupling model we can describe baryon asymmetry in low and high reheating temperature.

The paper is organized as follows:
In the section 2 we briefly introduce non-minimal derivative coupling model and obtain the field equations.
In the section 3 we examine conditions for oscillatory inflation and study the reheating phase in this model and the temperature at the end of warm inflation is calculated.
In the section 4 we discuss gravitational baryogenesis scenarios during oscillatory inflaton dominated in the context of non-minimal derivative coupling model and also we briefly investigate gravitational baryogenesis scenarios during radiation dominated phase.
In the section 5 we consider the qualitative implication of non-minimal derivative coupling by calculating the corresponding baryon asymmetry and compare our results with the observation.
In the last section we conclude our results.

We use units $\hbar=c=1$ through the paper.

\section{Brief review of non-minimal kinetic coupling model}

Let us consider the total action of non-minimal kinetic coupling model \cite{Germani,Germani2}

\begin{equation}\label{1.1}
S=\int \Big({M_P^2\over 2}R-{1\over 2}\Delta^{\mu \nu}\partial_\mu
\varphi \partial_{\nu} \varphi- V(\varphi)\Big) \sqrt{-g}d^4x+S_{int}+S_{r}+S_{B},
\end{equation}
where $\Delta^{\mu \nu}=g^{\mu \nu}+{1\over M^2}G^{\mu \nu}$, $G^{\mu \nu}=R^{\mu \nu}-{1\over 2}Rg^{\mu \nu}$ is the Einstein
tensor, $M$ is a coupling constant with mass dimension, $M_P=2.4\times 10^{18}GeV$ is the reduced Planck
mass, $S_{r}$ is the radiation action and $S_{int}$ describes the interaction of the scalar field with radiation.
In order to describe gravitational baryogenesis, we define action $S_B$ by interaction between derivative of Ricci scalar curvature $\partial_\mu R$ and baryon current $J^\mu$ as \cite{Davoudiasl}

 \begin{equation}\label{1.2}
S_{B}={1\over M_{*}^2}\int{d^4x\sqrt{-g}(\partial_\mu R)J^\mu},
\end{equation}
where $M_{*}$ is the cutoff scale of the effective theory.
We can obtain energy momentum tensor by variation of the action (\ref{1}) with respect to the metric,

\begin{equation}\label{2}
T_{\mu\nu}=T^{(\varphi)}_{\mu\nu}+T^{(r)}_{\mu\nu}.
\end{equation}
Where $T^{(r)}_{\mu\nu}$ is the energy momentum tensor for radiation described as

\begin{equation}\label{3}
T^{(r)}_{\mu\nu}=(\rho_r+P_r)u_{\mu}u_{\nu}+P_rg_{\mu\nu},
\end{equation}
where $u^{\mu}$ is the four-velocity of the radiation and $T^{(\varphi)}_{\mu\nu}$ is the energy momentum tensor for minimal and non-minimal
coupling counterparts of scalar field as
\begin{eqnarray}\label{4}
T^{(\varphi)}_{\mu\nu}&=&\nabla_{\mu}\varphi\nabla_{\nu}\varphi-{1\over2}g_{\mu\nu}{(\nabla\varphi)}^2-g_{\mu\nu}V(\varphi)\\\nonumber
&&-{1\over2}G_{\mu\nu}{(\nabla\varphi)}^2-{1\over2}R\nabla_{\mu}\varphi\nabla_{\nu}\varphi+
R^{\alpha}_{\mu}\nabla_{\alpha}\varphi\nabla_{\nu}\varphi\\ \nonumber
&&+R^{\alpha}_{\nu}\nabla_{\alpha}\varphi\nabla_{\mu}\varphi+R_{\mu\alpha\nu\beta}\nabla^{\alpha}\varphi\nabla^{\beta}\varphi
+\nabla_{\mu}\nabla^{\alpha}\varphi\nabla_{\nu}\nabla_{\alpha}\varphi\\\nonumber
&&-\nabla_{\mu}\nabla^{\nu}\varphi\Box \varphi
-{1\over2}g_{\mu\nu}\nabla^{\alpha}\nabla^{\beta}\varphi\nabla_{\alpha}\nabla_{\beta}\varphi+{1\over2}g_{\mu\nu}{(\Box\varphi)}^2\\\nonumber
&&-g_{\mu\nu}\nabla_{\alpha}\varphi\nabla_{\beta}\varphi R^{\alpha\beta}.
\end{eqnarray}

Energy transfer between the scalar field and radiation is assumed to be \cite{Berera,Berera2,Berera3,Berera4,Herrera1,Herrera2}
\begin{equation}\label{5}
Q_{\mu}=-\Gamma u^{\nu} \partial_{\mu}\varphi \partial_{\nu} \varphi.
\end{equation}

Where $\Gamma$ is decay rate of scalar field. In general, $\Gamma$ is a function of scalar field $\varphi$ and temperature $T$ \cite{Berera2,Xiao}.
The covariant derivative of energy momentum tensor becomes
\begin{equation}\label{6}
\nabla^{\mu}T^{(r)}_{\mu\nu}=Q_{\nu} \qquad and \qquad \nabla^{\mu}T^{(\varphi)}_{\mu\nu}=-Q_{\nu}.
\end{equation}

From relation (\ref{6}) for scalar field, the equation of motion in the spatially flat FLRW Universe and in presence of the dissipative term becomes
\begin{equation}\label{7}
\Big(1+{3H^2\over M^2}\Big)\ddot{\varphi}+3H\Big(1+{3H^2\over M^2}+{2\dot{H}\over M^2}\Big)\dot{\varphi}+V'(\varphi)+\Gamma\dot{\varphi}=0,
\end{equation}

where $H=\dot{a}/a$ is the Hubble parameter, overdot sign is derivative with respect to cosmic time $t$,
prime is derivative with respect to the scalar field $\varphi$ and $\Gamma\dot{\varphi}$ is the friction term which describes the decay of the scalar field to radiation. The Friedmann equations are given by
\begin{eqnarray}\label{8}
H^2&=&{1\over 3M_P^2}(\rho_\varphi+\rho_{r}),\\
\dot{H}&=&-{1\over 2M_P^2}(\rho_\varphi+\rho_{r}+P_\varphi+P_{r}).
\end{eqnarray}

Where $\rho_r$ and $P_r$ are the energy density and the pressure of radiation, respectively.
In the non-minimal derivative coupling model, the energy density $\rho_\varphi$ and the pressure of inflaton field $P_\varphi$ in FRW metric can be expressed as \cite{Germani,Germani2}
\begin{eqnarray}\label{9}
\rho_\varphi&=&(1+{9H^2\over M^2}){\dot{\varphi}^2\over 2}+V(\varphi),\\
P_\varphi&=&(1-{3H^2\over M^2}-{2\dot{H}\over M^2}){\dot{\varphi}^2\over 2}-V(\varphi)-{2H\dot{\varphi}\ddot{\varphi}\over M^2}.
\end{eqnarray}

Moreover, we can write energy density of radiation as a function of temperature $T$ and entropy density $s$, $\rho_{r}=(3/4)Ts$ \cite{Berera2}.
Using the equation of state parameter for radiation $\omega_r=1/3$, the evolution equation of energy density of radiation and scalar field becomes
\begin{eqnarray}\label{10}
\dot{\rho_{r}}+4H\rho_{r}&=&\Gamma{\dot{\varphi}}^2,\\
\dot{\rho_{\varphi}}+3H(\rho_{\varphi}+P_{\varphi})&=&-\Gamma{\dot{\varphi}}^2.
\end{eqnarray}

\section{Reheating after inflation}

\begin{figure}[t]
\begin{center}
  \scalebox{0.6}{\includegraphics{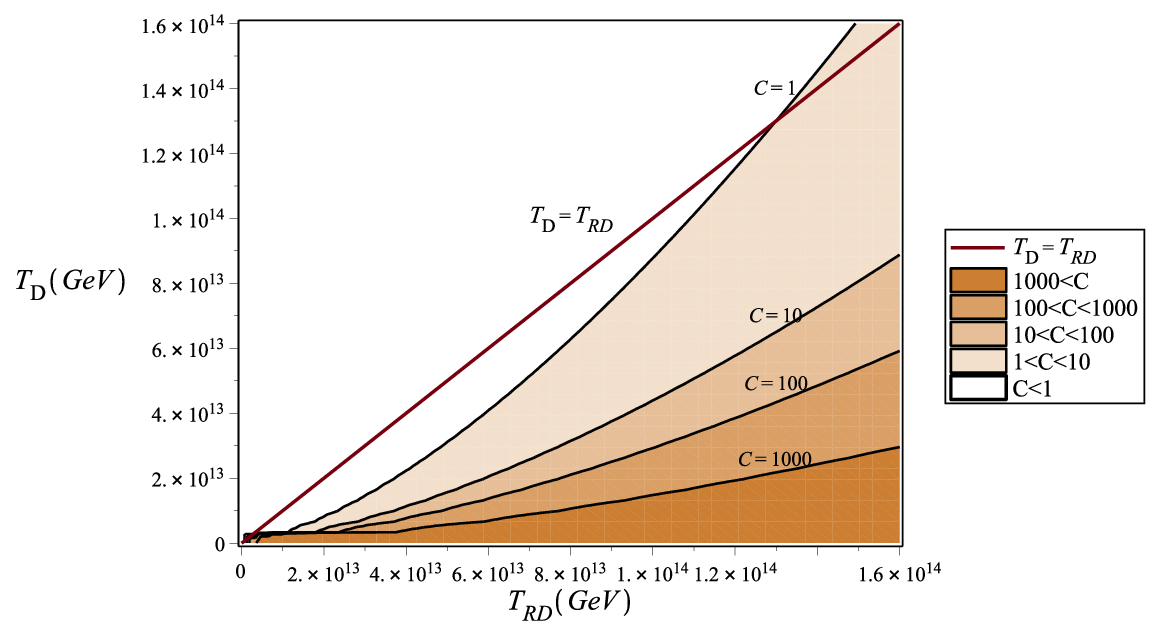}}
  \end{center}
   \caption{ \footnotesize The admissible region of decoupling temperature ${T_D}$ and reheating temperature $T_{RD}$ from high friction condition, in the case that decoupling take place during the reheating era for quadratic inflationary potential ($q=2$) and dimension-6 B-violating interaction ($n=2$).
   We assumes coupling constant $M=10^{-8}M_p$ and different values of $C$. The allowed regions $(C>10)$ are restricted to the dark brown.
    The bright colors of the region represent low values while the dark colors represent high values of $C$.}
\label{fig1}
\end{figure}

\begin{figure}[t]
\begin{center}
  \scalebox{0.6}{\includegraphics{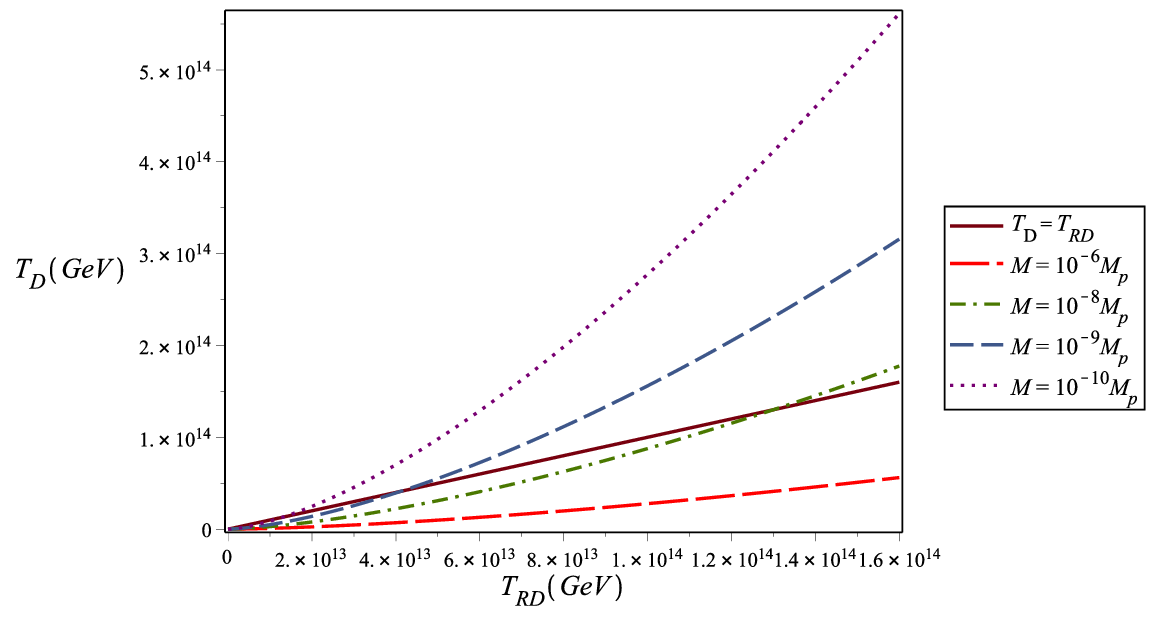}}
  \end{center}
   \caption{ \footnotesize Comparison high friction condition for different values of coupling constant $M$ from equation (\ref{42}). In the case that decoupling take place during the reheating era for quadratic inflationary potential ($q=2$), dimension-6 B-violating interaction ($n=2$) and $C=1$.}
\label{fig2}
\end{figure}

In this section we will consider reheating of the Universe after the end of slow-roll inflation. At the end of inflation, oscillation of the scalar field about the bottom of potential begins. We assume that the potential is even, $ V(-\varphi)=V(\varphi)$, and consider rapid oscillating solution to equation (\ref{7}) around $\varphi=0$. The inflaton energy density may estimated as $\rho_{\varphi}=V(\Phi(t))$, where $\Phi(t)$ is the amplitude of inflaton oscillation. In this epoch $\rho_\varphi$ and $H$ change insignificantly during a period of oscillation \cite{sadjadi1,sadjadi2}.

In the rapid oscillation of the scalar field epoch, the adiabatic index defined by $\gamma=\big<{\rho_\varphi+P_\varphi\over \rho_\varphi}\big>$, where bracket denotes the time average over an oscillation cycle.
Now, for a power law potential
\begin{equation}\label{11}
V(\varphi)=\lambda\varphi^q,
\end{equation}
at the high friction limit $(H^2/M^2\gg1)$, adiabatic index becomes \cite{sadjadi2}
\begin{equation}\label{12}
\gamma\approx{2q\over 3q+6}.
\end{equation}

Therefore, if we take the time average of both sides of the continuity equation (\ref{10}), we can write \cite{sadjadi3}
\begin{equation}\label{13}
{d\over dt}\big<\rho_\varphi\big>+3H\gamma\big<\rho_\varphi\big>+{\gamma\Gamma M^2 \over 3H^2}\big<\rho_\varphi\big>=0.
\end{equation}

We can simply derive the average of energy density of scalar field $\big<\rho_\varphi\big>$ at the high friction limit and $\Gamma M^2\ll 3H^3$ constrain as
\begin{equation}\label{14}
\big<\rho_\varphi\big>\propto a(t)^{-3\gamma}.
\end{equation}

By using relation (\ref{14}) and Friedmann equation (\ref{8}), in the scalar field dominated era, we can obtain scale factor as a function of the cosmic time
\begin{equation}\label{15}
a(t)\propto t^{{q+2\over q}}\propto t^{{2\over 3\gamma}}.
\end{equation}

Therefore, the Hubble parameter in the energy density of scalar field dominated era, can be estimated as $H\approx 2/(3\gamma t)$.
In the rapid oscillation phase, for the power law potential (\ref{11}), we can calculate the amplitude of the oscillations of scalar field as
\begin{equation}\label{16}
\phi(t)\propto a(t)^{-{2\over q+2}}\propto t^{-{2\over q}}.
\end{equation}

We have seen at high friction limit $\big<\dot{\varphi}^2\big>\approx\gamma M_P^2 M^2$ is nearly constant.
Therefore from equations (\ref{10}) and (\ref{11}) we can calculate evolution of radiation and scalar field energy density as
\begin{eqnarray}\label{17}
\rho_r&=&{3\Gamma \gamma^2 M^2 M_P^2\over (8+3\gamma)}t\bigg[1-({t_{o}\over t})^{(1+{8\over3\gamma})}\bigg],\\
\rho_{\varphi}&=&\rho_{o}\Big({t_{o}\over t}\Big)^2\exp{\Big[-\Big({\Gamma\gamma^3 M^2\over4}\Big)(t^3-t_{o}^3)\Big]},
\end{eqnarray}

where $t_{o}$ is the beginning of scalar field rapid oscillation, which $\rho_r(t=t_{o})=0$ and $\rho_{o}=\rho_{\varphi}(t_{o})$.
In the rapid oscillation phase, energy density of radiation increases slowly and energy density of scalar field decreases, so that at the time $t_{RD}$ the energy density of radiation becomes equal to the scalar field $\rho_r(t_{RD})\approx\rho_{\varphi}(t_{RD})$. From equations (\ref{17}) and (\ref{8}) we can calculate $t_{RD}$ as \cite{sadjadi2,sadjadi3}
\begin{equation}\label{18}
{t_{RD}}^3\approx{4(8+3\gamma)\over9\Gamma\gamma^4M^2}.
\end{equation}

We can write energy density of radiation as a function of reheating temperature $T_{RD}$ as \cite{Berera}
\begin{equation}\label{19}
\rho_r(t_{RD})=g_{\star}{\pi^2\over30}T_{RD}^4,
\end{equation}

where $g_{\star}$ is the number of degree of freedom at the reheating temperature and $T_{RD}$ is the temperature of radiation at the beginning of radiation dominated era. Therefore the temperature of the universe at the beginning of radiation dominated universe becomes
\begin{equation}\label{20}
{T_{RD}}^4\approx {30M_P^2\over\pi^2g_{\star}}\bigg[{12\Gamma^2\gamma^2M^4\over{(8+3\gamma)}^2} \bigg]^{1\over3}.
\end{equation}

\section{Gravitational baryogenesis}

In order to examine the effect of non-minimal derivative coupling and the high friction condition on the baryon asymmetry, in the present work,
we consider the baryon number-to-entropy ratio during reheating era and radiation dominated period.

\begin{figure}[t]
\begin{center}
  \scalebox{0.6}{\includegraphics{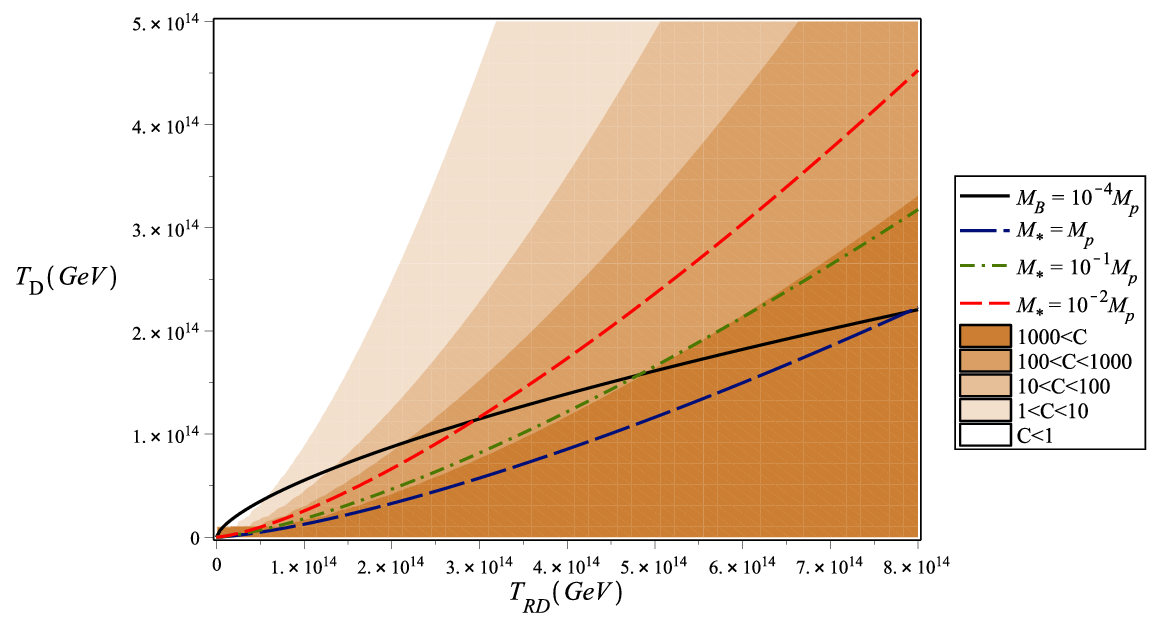}}
  \end{center}
   \caption{ \footnotesize Decoupling temperature ${T_D}$ in terms of reheating temperature $T_{RD}$, for coupling constant $M=10^{-8}M_p$ and high reheating temperature. The blue, green and red dashed curves correspond to different values of cutoff scale $M_\star$, to explain the observed baryon asymmetry ($Y_B=8.64\times 10^{-11}$). The solid black curve correspond to “decoupling of B-violating processes” with $M_B=10^{-4}M_p$.
   In the case that decoupling take place during the reheating era for quadratic inflationary potential ($q=2$) and dimension-6 B-violating interaction ($n=2$).The admissible regions for decoupling temperature and reheating temperature are displayed with a brown color spectrum, from high friction condition. Intersection of the dashed curves and solid curve are the points where defines, $T_{RD}$ and ${T_D}$.}
\label{fig3}
\end{figure}

\begin{figure}[t]
\begin{center}
  \scalebox{0.6}{\includegraphics{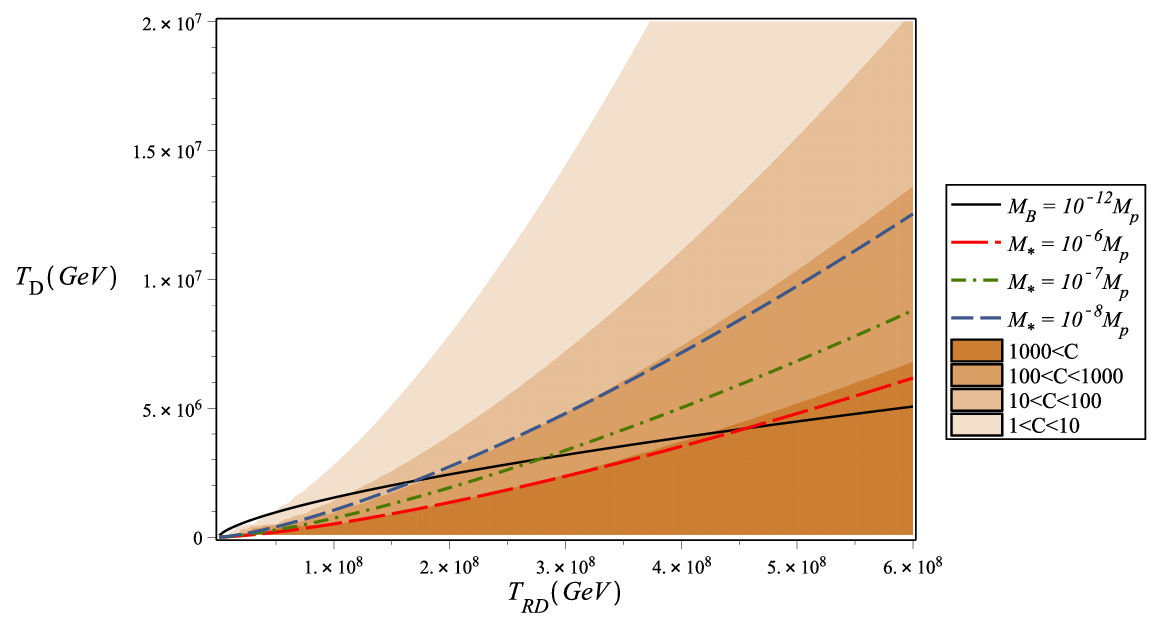}}
  \end{center}
   \caption{ \footnotesize Decoupling temperature ${T_D}$ in terms of reheating temperature $T_{RD}$, for coupling constant $M=10^{-14}M_p$ and low reheating temperature. The blue, green and red dashed curves correspond to different values of cutoff scale $M_\star$, to explain the observed baryon asymmetry ($Y_B=8.64\times 10^{-11}$) in relation (\ref{43}). The solid black curve correspond to “decoupling of B-violating processes” with $M_B=10^{-12}M_p$. In the case that decoupling take place during the reheating era for quadratic inflationary potential ($q=2$) and dimension-6 B-violating interaction ($n=2$). The admissible regions for decoupling temperature and reheating temperature are displayed with a brown color spectrum, from high friction condition. Intersection of the dashed curves and solid curve are the points where defines, $T_{RD}$ and ${T_D}$.}
\label{fig4}
\end{figure}

\subsection{Reheating period}

During the reheating era, the scalar field oscillates about minimum of potential and decays to ultra relativistic particles. In this period the energy density of oscillatory scalar field is dominated and the Universe expansion is accelerated.
In this period we can rewrite the equation (\ref{15}) in the form of

\begin{equation}\label{21}
a(t)=a_{RD}\bigg({t\over t_{RD}}\bigg)^{2\over3\gamma},
\end{equation}

where $a_{RD}$ is scale factor at the beginning of the radiation dominated era $t_{RD}$.
The evolution of the energy density of radiation during the rapid oscillation of scalar field from relation (\ref{10}) becomes

\begin{equation}\label{22}
{d\over dt}(a^4\rho_{r})\approx -3\Gamma\gamma M^2 M_P^2 a^4 \Rightarrow \rho_r \propto a^{3\gamma\over2}.
\end{equation}

Therefore we can write the energy density of radiation as a function of scale factor as
\begin{equation}\label{23}
\rho_r={g_{\star}\pi^2\over30}T_{RD}^4 \bigg({a\over a_{RD}}\bigg)^{3\gamma\over2},
\end{equation}
where $(g_{\star}\pi^2/30)T_{RD}^4$ is energy density of radiation at the $t_{RD}$.
From relation (\ref{14}) and the radiation-scalar field equality $\rho_\varphi(t_{RD})=\rho_r(t_{RD})$,
 the energy density of scalar field is given by
\begin{equation}\label{24}
\rho_\varphi= {g_{\star}\pi^2\over30}T_{RD}^4 \bigg({a\over a_{RD}}\bigg)^{-3\gamma}.
\end{equation}

From relation (\ref{23}) and Fridmann equation (\ref{8}), during reheating period, we can calculate evolution of scale factor as a function of temperature as
\begin{equation}\label{25}
a=a_{RD} \bigg({T\over T_{RD}}\bigg)^{8\over3\gamma}.
\end{equation}
This relation show that in the non-minimal derivative coupling model, the temperature of Universe increases by expansion of the Universe until beginning of radiation dominated period. While in the standard thermal history of the Universe, temperature had opposite evolution.
By replacement of relation (\ref{25}) into equation (\ref{24}) the energy density of inflaton field becomes
\begin{equation}\label{26}
\rho_\varphi= g_{\star}{\pi^2\over30}\bigg({T_{RD}^{12}\over T^8}\bigg).
\end{equation}

If there exist B-violating interaction in thermal equilibrium then it can generate net baryon asymmetry.
In the expanding Universe From action (\ref{1.2}) we have \cite{Davoudiasl,Fukushima}

\begin{equation}\label{27}
 {1\over M_\star}(\partial_\mu)J^\mu={\dot{R}\over M_{\star}^2}(g_bn_b+g_{\bar{b}}n_{\bar{b}}),
\end{equation}

where $g_b=-g_{\bar{b}}$  denotes the number of intrinsic degree of freedom of baryons. $n_b$ and $n_{\bar{b}}$ are the number densities of baryon and antibaryon respectively. An effective chemical potential follow as $\mu_b=-\mu_{\bar{b}}=g_b\dot{R}/M_{\star}^2$, the entropy density of the Universe is given by $s=2\pi^2g_\star T^3/45$, the baryon number density, in thermal equilibrium, becomes $n_B=(g_bn_b+g_{\bar{b}}n_{\bar{b}})=-g_b\mu_bT^2/6$ \cite{kolb}.
As a result, we can write the baryon to entropy ratio (baryon asymmetry) in an accelerating universe as
\begin{equation}\label{28}
Y_B\equiv {n_B\over s}\approx -{15g_b^2\over4\pi^2 g_\star}{\dot{R}\over M_{\star}^2T}\Big|_{T=T_{D}},
\end{equation}
where temperature $T_D$ is the temperature of the Universe at which the baryon current violation decouples.
In the spatially flat FLRW metric, of the form,
\begin{equation}\label{28.1}
ds^2=-dt^2+a(t)^2\sum(dx^i)^2,
\end{equation}
Ricci scalar curvature $R$ is equal to
\begin{equation}\label{29}
R=-6(\dot{H}+2H^2),
\end{equation}
by using equation (\ref{25}) for the scale factor, easily we can calculate Ricci scalar curvature as a function of cosmic time as
\begin{equation}\label{30}
R=4\Big({4-3\gamma\over3\gamma^2} \Big)t^{-2}.
\end{equation}
Now, with the time derivative of Ricci scalar curvature (\ref{30}) and Fridmann equation $H^2\approx \rho_\varphi/(3M_p^2)$ in the scalar field dominated period, we have
\begin{equation}\label{31}
\dot{R}\approx -\sqrt{3}\gamma(4-3\gamma){\rho_{\varphi}^{3\over2}\over M_p^3}.
\end{equation}

Finally, by substituting $\dot{R}$ and $\rho_\varphi$ from equations (\ref{31}) and (\ref{26}) into equation (\ref{28}), we obtain baryon asymmetry $Y_B$ as a function of Universe temperature
\begin{equation}\label{32}
Y_B\equiv {n_B\over s}\approx  {\pi\gamma(4-3\gamma) g_b^2 \sqrt{g_\star}\over 8\sqrt{10}}
{T_{RD}^{18}\over M_p^3 M_\star^2 T^{13}}\Big|_{T=T_{D}}.
\end{equation}

We continue with a brief mention of the origin of the B-violating interaction that is indispensable for any baryogenesis scenario. We assume that  B-violating interactions, which are given by an operator $\mathcal{O}_B$ of mass dimension $D=4+n$ \cite{Davoudiasl,Fukushima}. We need $n>0$ for the B-violating interaction. In the B-violating interactions, coupling constants are proportional to $M_B^{-n}$, where $M_B$ is the mass scale, the rate of generation of B-violating interaction in thermal equilibrium with the temperature $T$ can be cast in the form \cite{Davoudiasl}
\begin{equation}\label{33}
\Gamma_B={T^{2n+1}\over M_B^{2n}}.
\end{equation}

Decoupling of B-violating processes occurs at $T=T_D$, when $\Gamma$ falls below $H={2/3\gamma t}$.
Therefore we can obtain temperature of decoupling from equation (\ref{26}) as
\begin{equation}\label{34}
T_D=\Bigg({\pi\sqrt{g_{\star}}M_B^{2n}\over3\sqrt{10} M_p}\Bigg)^{1\over2n+5}T_{RD}^{6\over2n+5}.
\end{equation}

Therefore by substituting of relation (\ref{34}) into equation (\ref{32}) we have
\begin{equation}\label{35}
Y_B\approx\Big({3\gamma(4-3\gamma)g_b^2\over 8}\Big)\Big({\pi^2g_{\star}\over 90}\Big)^{\big({n-4\over2n+5}\big)}
{T_{RD}^{12\big({3n+1\over2n+5}\big)} \over M_\star^2 M_p^{2\big({3n+1\over2n+5}\big)} M_B^{\big({26n\over2n+5}\big)}}.
\end{equation}

From high friction condition $H^2/M^2\gg1$ we can constrain the decoupling temperatures $T_D$ as
\begin{equation}\label{36}
MM_pT_D^4\ll \sqrt{{g_{\star}\pi^2\over90}}T_{RD}^6.
\end{equation}

Contrary to minimal coupling model, we have seen during the oscillatory scalar field dominated universe for $\gamma=1/3$ the temperature of the universe increases as $T\approx T_{RD}\big(a(t)/a_{RD}\big)^{(1/8)}$.
Therefore, the decoupling temperature $T_D$ is smaller than the reheating temperature $T_{RD}$, in pleasant accordance with the condition (\ref{36}).

\begin{figure}[t]
\begin{center}
  \scalebox{0.6}{\includegraphics{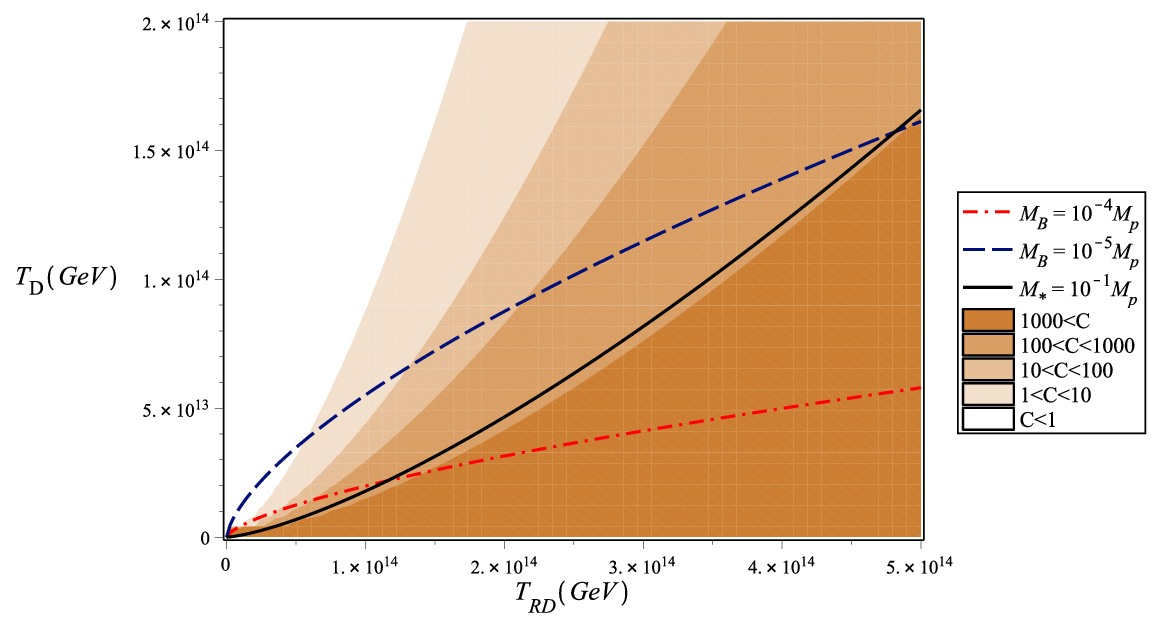}}
   \end{center}
    \caption{ \footnotesize Decoupling temperature ${T_D}$ in terms of reheating temperature $T_{RD}$, for coupling constant $M=10^{-8}M_p$ and high reheating temperature. The blue and red dashed curves correspond to different values of $M_B$ in "decoupling of B-violating processes".
    The solid black curve correspond to cutoff scale $M_\star=10^{-1}M_p$, to explain the observed baryon asymmetry ($Y_B=8.64\times 10^{-11}$).
   In the case that decoupling take place during the reheating era for quadratic inflationary potential ($q=2$) and dimension-6 B-violating interaction ($n=2$). The admissible regions for decoupling temperature and reheating temperature are displayed with a brown color spectrum, from high friction condition. Intersection of the dashed curves and solid curve are the points where defines, $T_{RD}$ and ${T_D}$.}
    \label{fig5}
\end{figure}

\begin{figure}[t]
\begin{center}
  \scalebox{0.6}{\includegraphics{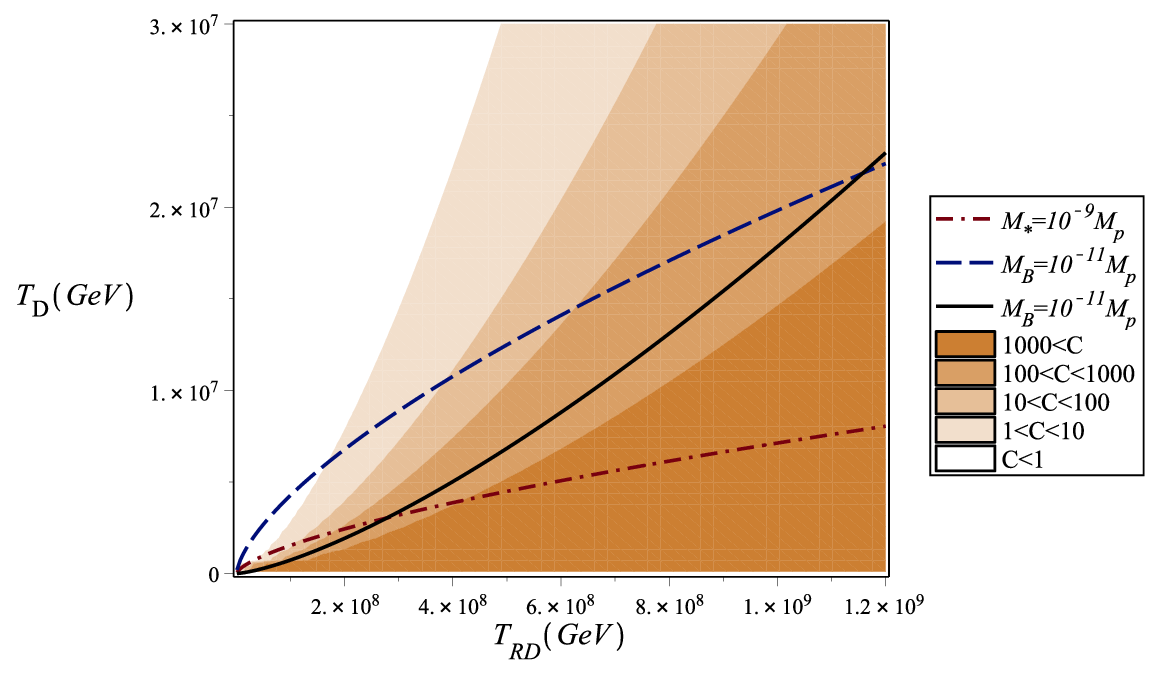}}
  \end{center}
   \caption{ \footnotesize Decoupling temperature ${T_D}$ in terms of reheating temperature $T_{RD}$, for coupling constant $M=10^{-14}M_p$ and low reheating temperature. The blue and red dashed curves correspond to different values of $M_B$ in "decoupling of B-violating processes".
    The solid black curve correspond to cutoff scale $M_\star=10^{-9}M_p$, to explain the observed baryon asymmetry ($Y_B=8.64\times 10^{-11}$).
   In the case that decoupling take place during the reheating era for quadratic inflationary potential ($q=2$) and dimension-6 B-violating interaction ($n=2$). The admissible regions for decoupling temperature and reheating temperature are displayed with a brown color spectrum, from high friction condition. Intersection of the dashed curves and solid curve are the points where defines, $T_{RD}$ and ${T_D}$.}
     \label{fig6}
\end{figure}

\subsection{Radiation dominated period}

In this section we consider the gravitational baryogenesis during the radiation dominated era after the reheating period.
The challenge of gravitational baryogenesis during radiation dominated universe is characterised by equation of state $\omega\approx1/3$.
If $\omega$ were equal to $1/3$ exactly, then $R=3(1-3\omega)H^2$ would vanish and $Y_B=0$. Then the baryon asymmetry effect would never be generated during the radiation dominated era.
If we look more closely at the problem, it is not so serious, so that if we take into account the effect of interactions among massless particles lead to trace anomaly that make $T_\mu^\mu\neq0$. Therefore the equation of state is given by $1-3\omega\sim10^{-2}-10^{-1}$ \cite{Davoudiasl,Fukushima}.

In the radiation dominated era $a\propto\sqrt{t}$, $\rho_r\propto a^{-4}$, $H\approx1/2t$ and by relation $\rho_r=(\pi^2g_{\star}/30)T^4$ we arrive at

\begin{equation}\label{37}
T^2={3\sqrt{10}M_p\over 2\pi\sqrt{g_{\star}}}t^{-1}.
\end{equation}

We can calculate the time of decoupling by equality $\Gamma=H$, as $t_D\approx M_B^{2n}/2T_D^{2n+1}$, and the temperature of decoupling becomes
\begin{equation}\label{38}
T_D^{2n-1}\approx {\pi M_B^{2n}\sqrt{g_{\star}}\over 3\sqrt{10}M_p}.
\end{equation}
By using Fridmann equation during of radiation dominated era $H^2\approx\rho_r/3M_p^2$ we have $R=3(1-3\omega)(1/4t^2)$, therefore baryon asymmetry reads

\begin{equation}\label{39}
Y_B\approx {g_b^2\over2}\Big({\pi^2g_{\star}\over90}\Big)^{\big({n+2\over2n-1}\big)}(1-3\omega) M_{\star}^{-2} M_p^{-\big({6n+2\over2n-1}\big)} M_B^{\big({10n\over2n-1}\big)}.
\end{equation}
Hence, by choosing dimension-6 B-violating interaction ($n=2$) and $g_{\star}=106$ we have
\begin{equation}\label{40}
Y_B\approx 13(1-3\omega) M_{\star}^{-2} M_p^{-\big({14\over3}\big)} M_B^{\big({20\over3}\big)}.
\end{equation}

Similarly, by choosing dimension-5 B-violating interaction ($n=1$) we have
\begin{equation}\label{41}
Y_B\approx 785(1-3\omega) M_{\star}^{-2} M_p^{-8} M_B^{10}.
\end{equation}

\section{numerical analysis}

\begin{figure}[t]
\begin{center}
  \scalebox{0.6}{\includegraphics{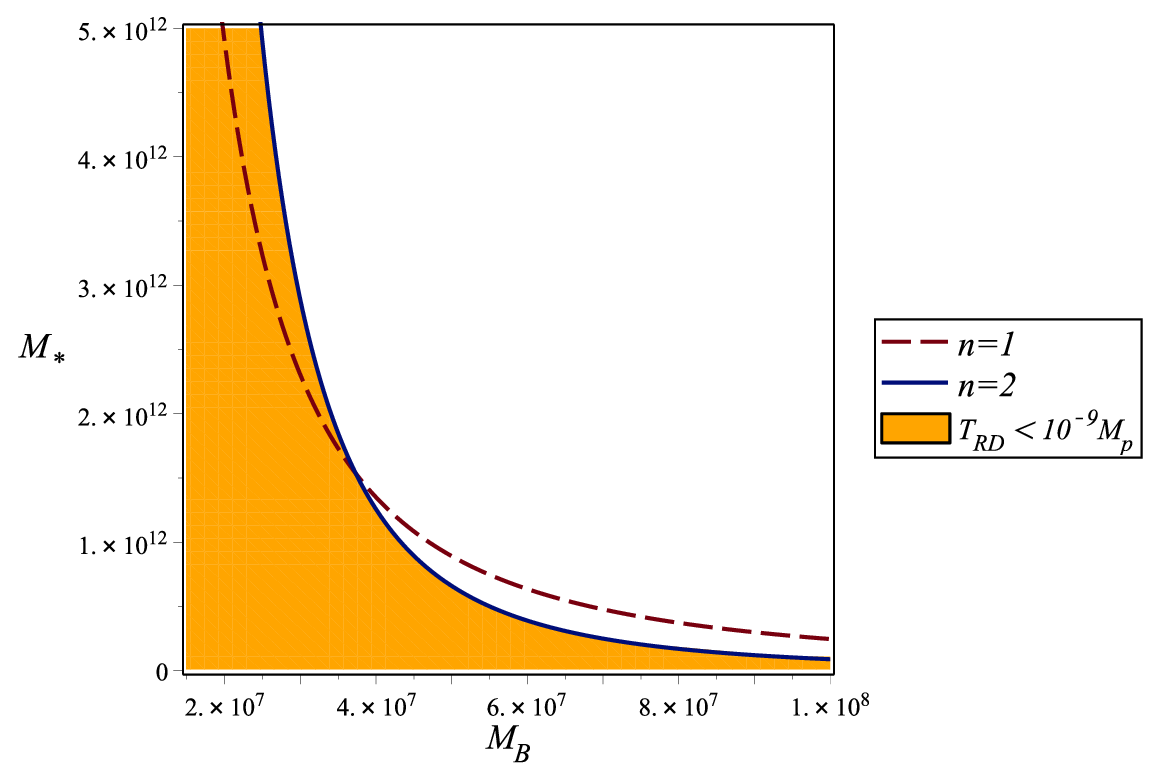}}
  \end{center}
    \caption{ \footnotesize The acceptable range for $M_\star$ and $M_B$ to explain the observed baryon asymmetry ($Y_B=8.64\times 10^{-11}$).
The solid blue curve, correspond to dimension-6 B-violating interaction ($n=2$) and red dashed curve dimension-5 B-violating interaction ($n=1$).
We assume that reheating temperature is $T_{RD}=10^{-9}M_p$ and the golden region show the $T_{RD}<10^{-9}M_p$.}
  \label{fig7}
\end{figure}
As we have shown in the previous section, the value of reheating temperature $T_{RD}$ is depends on coupling constant $M$.
On the other hand, high friction condition, constrain the reheating temperature.
To display high friction condition $H^2/M^2\gg1$ in our plots, we can parameterize relation (\ref{36}) as
\begin{equation}\label{42}
1\ll C\equiv\sqrt{{g_{\star}\pi^2\over90}}{T_{RD}^6\over MM_pT_D^4}.
\end{equation}

As we can see, this constraint not only does not depend on $q$ and $n$ parameters, but also it depends only on coupling constant $M$.
Therefore, for any value of coupling constant $M$, we must chose $T_D$ and $T_{RD}$ where satisfy in this condition.
In all of the plots Figure \ref{fig1}-Figure \ref{fig6}, decoupling take place during the reheating era for quadratic inflationary potential ($q=2$ or $\gamma=1/3$). Also, we assume that dimension-6 B-violating interaction ($n=2$), the ultra relativistic degrees of freedom at the electroweak energy scale $g_{\star}=106.75$, the number of intrinsic degree of freedom of baryons $g_b\approx\mathcal{O}(1)$.

In Figure \ref{fig1} the admissible region of decoupling temperature ${T_D}$ and reheating temperature $T_{RD}$ from high friction condition, in the case that decoupling take place during the reheating era for quadratic inflationary potential ($q=2$), dimension-6 B-violating interaction ($n=2$), coupling constant $M=10^{-8}M_p$, and different values of $C$ has been depicted. As a reference, the acceptable regions are $C>10$ where displayed by the brown color. So the region $C<1$ is out of bounds and the values of reheating temperature and decoupling temperature in this region violate the high friction condition. The bright colors represent low values while the dark colors represent high values for $C$.

We can rewrite relation (\ref{32}), (\ref{34}) and (\ref{35}) for $\gamma=1/3$ and $n=2$, in the form of
\begin{equation}\label{43}
Y_B\approx{3\over8}\Bigg({\pi\sqrt{g_\star}\over 3\sqrt{10}}\Bigg){T_{RD}^{18}\over M_p^3 M_\star^2 T_D^{13}},
\end{equation}

\begin{equation}\label{44}
T_D\approx\Bigg({\pi\sqrt{g_\star}\over 3\sqrt{10}}\Bigg)^{1/9}{M_{B}^{4/9}\over M_p^{1/9}}T_{RD}^{2/3},
\end{equation}

\begin{equation}\label{45}
Y_B\approx {3\over8}\Bigg({\pi\sqrt{g_\star}\over 3\sqrt{10}}\Bigg)^{(-4/9)}{T_{RD}^{28/3}\over M_{\star}^{2} M_p^{14/9} M_B^{52/9}}.
\end{equation}

In Figure \ref{fig2} high friction condition (\ref{42}) for different values of coupling constant $M$ has been depicted. Clearly, as the coupling constant $M$ is decreased, the acceptable region for $T_{RD}$ and $T_D$ associated with high temperature and vice versa.

Figure \ref{fig3} shows decoupling temperature ${T_D}$ in terms of reheating temperature $T_{RD}$, for coupling constant $M=10^{-8}M_p$.
The blue, green and red dashed curves correspond to different values of cutoff scale $M_\star$, to explain the observed baryon asymmetry ($Y_B=8.64\times 10^{-11}$) in relation (\ref{43}). The solid black curve correspond to "decoupling of B-violating processes" with $M_B=10^{-4}M_p$ in relation (\ref{44}).
The admissible regions for decoupling temperature and reheating temperature are displayed with a brown color spectrum, from high friction condition.
Intersection of the dashed curves and solid curve are the points where defines, $T_{RD}$ and ${T_D}$.
We have seen, intersection points take place in admissible regions $C>100$ and $C>1000$.
Therefore, if we want to explain, generation of baryon asymmetry in high reheating temperature, then we have to choose higher values of $M_\star$ and $M_B$.

Therefore more appropriate choices, are larger cutoff scale $M_\star$ which fall into the dark  eras.

The Figure \ref{fig4} is similar to Figure \ref{fig3}, but we drawing this plot for low reheating temperature and different parameters values.
We have seen, if we want to explain, generation of baryon asymmetry in low reheating temperature, then we have to choose smaller values of $M_\star$, $M_B$ and $M$.

The Figure \ref{fig5} shows decoupling temperature ${T_D}$ in terms of reheating temperature $T_{RD}$, for coupling constant $M=10^{-8}M_p$ and high reheating temperature. The blue and red dashed curves correspond to different values of $M_B$ in "decoupling of B-violating processes" in relation (\ref{44}). The solid black curve correspond to cutoff scale $M_\star=10^{-1}M_p$, to explain the observed baryon asymmetry ($Y_B=8.64\times 10^{-11}$) in relation (\ref{43}). The admissible regions for decoupling temperature and reheating temperature are displayed with a brown color spectrum, from high friction condition. Intersection of the dashed curves and solid curve are the points where defines, $T_{RD}$ and ${T_D}$ where satisfy in relations (\ref{43}), (\ref{44}) and high friction condition. We have seen, intersection points take place in admissible regions $C>100$.

The Figure \ref{fig6} is similar to Figure \ref{fig5}, but we drawing this plot for low reheating temperature and different parameters values.
We have seen, if we want to explain, generation of baryon asymmetry in low reheating temperature, then we have to choose smaller values of $M_\star$, $M_B$ and $M$.

The Figure \ref{fig7}, the acceptable range for $M_\star$ and $M_B$ to explain the observed baryon asymmetry ($Y_B=8.64\times 10^{-11}$) has been depicted.
The solid blue curve, correspond to dimension-6 B-violating interaction ($n=2$) and red dashed curve dimension-5 B-violating interaction ($n=1$).
We assume that reheating temperature is $T_{RD}=10^{-9}M_p$ and the golden region show the $T_{RD}<10^{-9}M_p$.

We conclude that in non-minimal derivative coupling model, sufficient baryon asymmetry has been generated in low and high reheating temperature during reheating phase.

\section{Conclusion}
In this paper, We investigated the gravitational baryogenesis mechanism in the non-minimal derivative coupling model in high friction regime.
We used coupling between derivative of Ricci scalar curvature and baryon current to describe baryon asymmetry.
In this model, inflaton begins a coherent rapid oscillation, after the slow roll inflation. During this stage, inflaton decays to radiation and reheats the Universe.
We calculated the baryon to entropy ratio in the case that reheating period described by coherent rapid oscillation in the non-minimal derivative coupling model.
Moreover this model has the additional constraint, where we have to choose the parameters of the model in such a way that it satisfies this additional constraint. We examine this condition in all of plots.
As we demonstrated, in contrast to the standard gravitation baryogenesis where could not explain baryon asymmetry in low reheating temperature, in the non-minimal derivative coupling model we can describe baryon asymmetry in low and high reheating temperature.


\begin{thebibliography}{99}

\bibitem {WMAP} C. L. Bennett et al. [WMAP Collaboration], Astrophys. J. Suppl. 148, 1 (2003).
\bibitem {Cohen} Andrew G. Cohen, A. De Rujula, S.L. Glashow, Astrophys. J. 495, 539 (1998).
\bibitem {BBN} Cyburt, Richard H. et al. Rev.Mod.Phys. 88 (2016) 015004.
\bibitem {Planck 2015} Planck 2015 results. XIII. Cosmological parameters - Planck Collaboration (Ade, P.A.R. et al.) Astron.Astrophys. 594 (2016) A13.
\bibitem {Sakharov} A. D. Sakharov, JETP Lett. 5, 24 (1967).
\bibitem {Weinberg} S. Weinberg, Phys. Rev. Lett. 42, 850 (1979).
\bibitem {Affleck} I. Affleck and M. Dine, Nucl. Phys. B249, 361 (1985).
\bibitem {Fukugita} M. Fukugita and T. Yanagida, Phys. Lett. B174, 45 (1986).
\bibitem {Cohen1} A. Cohen, D. Kaplan, 1987, Phys. Lett. B, 199, 251.
\bibitem {Cohen2} A. Cohen, D. Kaplan, A.E. Nelson, 1991, Phys. Lett. B, 263, 86.
\bibitem {Davoudiasl} H. Davoudiasl, R. Kitano, G. D. Kribs, H. Murayama and P. J. Steinhardt, Phys. Rev. Lett. 93, 201301 (2004).
\bibitem {Sadjadi} H. Mohseni Sadjadi,:Phys. Rev. D 76, 123507 (2007).
\bibitem {Lambiase} G. Lambiase, G. Scarpetta, Phys. Rev. D 74, 087504 (2006).
\bibitem {Ramos} M. P. L. P. Ramos and J. Paramos, Phys. Rev. D 96, 104024 (2017).
\bibitem {Nojiri} S. Nojiri, S. D. Odintsov, Int. J. Geom. Meth. Mod. Phys. 4 (2007) 115.
\bibitem {Capozziello} Salvatore Capozziello, Mariafelicia De Laurentis, Physics Reports, 509, 167, (2011).
\bibitem {Capozziello1} Salvatore Capozziello, Rocco D’Agostino, Orlando Luongo, Int.J.Mod.Phys.D 28 (2019) 10, 1930016.
\bibitem {Saridakis} V. K. Oikonomou and Emmanuel N. Saridakis, Phys. Rev. D 94, 124005, (2016).
\bibitem {Odintsov} S.D. Odintsov, V.K. Oikonomou, Phys. Lett. B 760 (2016) 259-262.
\bibitem {Baffou} E. H. Baffou, M. J. S. Houndjo, D. A. Kanfon and I. G. Salako, Eur. Phys. J. C 79 (2019) 112.
\bibitem {Sahoo} P.K. Sahoo, Snehasish Bhattacharjee, Int. J. Theor. Phys. 59 (2020) 1451.
\bibitem {Bhattacharjee} Snehasish Bhattacharjee, P.K. Sahoo, Eur.Phys.J.C 80 (2020) 3, 289.
\bibitem {Azhar} Nadeem Azhar, Abdul Jawad, Shamaila Rani, Phys.Dark Univ. 30 (2020) 100724.
\bibitem {Nozari} Kourosh Nozari, Fateme Rajabi,Commun.Theor.Phys. 70 (2018) 4, 451.
\bibitem {Lima} J.A.S. Lima, D. Singleton, Physics Letters B 762 (2016) 506–511.
\bibitem {Saaidi} Kh. Saaidi and H. Hossienkhani, Astrophys. Space Sci. 333, 305 (2011).
\bibitem {Saaidi2} A. Aghamohammadi, H. Hossienkhani, Kh. Saaidi, Mod.Phys.Lett.A 33 (2018) 13, 1850072.
\bibitem {Fukushima} Mitsuhiro Fukushima, Shuntaro Mizuno, Kei-ichi Maeda, Phys. Rev. D 93, 103513 (2016).
\bibitem {Germani} C. Germani and A. Kehagias, Phys. Rev. Lett. 105 (2010) 011302.
\bibitem {Germani2} C. Germani and A. Kehagias, JCAP 05 (2010) 019.
\bibitem {Germani3} C. Germani and Y. Watanabe, JCAP 07 (2011) 031.
\bibitem {Horndeski} G.W. Horndeski, Int. J. Theor. Phys. 10 (1974) 363.
\bibitem {Tsutomu} Tsutomu Kobayashi, Rep. Prog. Phys. 82, 086901 (2019).
\bibitem {Nojiri2} Shin'ichi Nojiri, Sergei D. Odintsov, Phys. Rept. 505 (2011) 59-144.
\bibitem {Gao} X. Gao, Phys. Rev. D 90 (2014) 081501.

\bibitem {Amendola} L. Amendola, Phys. Lett. B 301, 175 (1993).
\bibitem {Charmousis} C. Charmousis, E.J. Copeland, A. Padilla and P.M. Saffin, Phys. Rev. Lett. 108 (2012) 051101.

\bibitem {sadjadi} H. M. Sadjadi, P. Goodarzi, Phys. Lett. B 732, 278 (2014).
\bibitem {sadjadi1} H. M. Sadjadi, P. Goodarzi , JCAP 02, 038, (2013).
\bibitem {sadjadi2} H. M. Sadjadi, P. Goodarzi, JCAP 07, 039 (2013).
\bibitem {nozari} K. Nozari, M. Shoukrani, and  N. Rashidi, Adv. High Energy Phys. 2014, 343819 (2014);
\bibitem {sadjadi3} H. M. Sadjadi, P. Goodarzi,  Eur. Phys. J. C 75, 513 (2015);
\bibitem {Berera} A. Berera, Phys. Rev.Lett. 75,3218,(1995); A. Berera, L. Z. Fang, Phys. Rev. Lett. 74,1912,(1995); Y. Gim and W. Kim, arXiv:1608.07466.
\bibitem {Berera2} A. Berera, Nucl. Phys. B 585, 666 (2000).
\bibitem {Berera3} L. M. H. Hall, I.G. Moss, and A. Berera, Phys. Rev. D 69,083525 (2004).
\bibitem {Berera4} M. Bastero-Gil, A. Berera, R. O. Ramos and J. G. Rosa, Phys. Lett. B 712, 425 (2012).
\bibitem {Herrera1} S. del Campo, R. Herrera and D. Pavon, Phys. Rev. D 75, 083518 (2007).
\bibitem {Herrera2} R. Herrera, S. del Campo and C. Campuzano, JCAP 10, 009 (2006).
\bibitem {Xiao} X. M. Zhang, J. Y. Zhu, Phys. Rev. D 87.043522(2013).
\bibitem {kolb} E. Kolb, M.Turner, The Early Universe (Addison-Wesley Publishing Company, Redwood City, California, 1990).
\bibitem {guth} A. H. Guth, Phys. Rev. D 23, 347 (1981).
\bibitem {inflaton1} A. Linde, Particle Physics and Inflationary Cosmology (Harwood, Chur, Switzerland, 1990).
\bibitem {Liddle} A. Liddle and A. Mazumder, Phys. Rev. D 58, 083508 (1996).
\bibitem {Zhang} X. M. Zhang,  J. Y. Zhu, Phys. Rev. D 87, 043522 (2013).
\end{thebibliography}
\end{document}